 \documentclass[final,5p,times,twocolumn]{elsarticle}

\usepackage{amssymb}
\usepackage{lipsum}
\usepackage{url}
\biboptions{numbers,sort&compress}

\begin{document}

\begin{frontmatter}

\title{Understanding Charge Radii with Machine Learning: Discovering Physics Expressions}

\author[first]{B. Maheshwari}
\author[first]{P. Van Isacker}
\affiliation[first]{organization={Grand Accélérateur National d'Ions Lourds}, address={CEA/DSM-CNRS/IN2P3}, street={Bvd Henri Becquerel}, pin={F-14076}, city={ Caen}, country={France}}

\begin{abstract}

We introduce a robust, interpretable machine learning (ML) framework that combines numerical regression for high-accuracy predictions with symbolic regression to uncover the underlying physics. This hybrid approach effciently derives analytical expressions by leveraging the smoothed predictions of optimized ML models, a significant acceleration over direct symbolic regression on raw experimental data. 

We apply this framework, as an example, to nuclear charge radii across the nuclear chart, notably including light nuclei that are often excluded from such studies. We employ Light Gradient Boosting Machine (LGBM) and Gaussian Process Regression (GPR) models to map correlations between charge radii and key physical features: mass $A^{1/3}$ and proton number $Z^{1/3}$ dependencies, total binding energy, and for the first time, the pairing gap. Our models are rigorously trained using four-fold cross-validation with automated hyperparameter optimization, ensuring robustness and generalizability, which is critical for the typically small and skewed datasets in nuclear physics. Finally, we distill the knowledge from the initial LGBM and GPR models into simplified, interpretable mathematical expressions via symbolic regression, white-boxing these ML models. The derived formulas provide physical insights comparable to traditional many-body models and demonstrate a powerful pathway for physics expression discovery guided by ML. 
\end{abstract}

\end{frontmatter}

\section{Introduction}

The evolution of nuclear charge radii across the nuclear chart is central to unraveling the coupling of nucleons and their interactions~\cite{Wood1992,Bender2003,Geithner2008,Wienholtz2013,Marsh2018,Groote2020,Yang2023}. This led to the development and application of various nuclear models of charge radii, ranging from macroscopic semi-empirical approaches such as the liquid drop model~\cite{Weizsacker1935,Duflo1994}, Garvey-Kelson relation~\cite{{Piekarewicz2010}}, Weizs\"acker-Skyrme model~\cite{Wang2013}, interacting boson model (IBM)~\cite{Zerguine2012}, to microscopic frameworks. Some examples of the latter include  Hartee-Fock-Bogoliubov (HFB)~\cite{Goriely2013,Stoitsov2003}, relativistic mean-field (RMF) theory~\cite{Niksic2015, Niksic2017}, and density-functional theory (DFT)~\cite{Meng2016}. Fundamental descriptions based on ab-initio methods have also been explored for lighter nuclei~\cite{Shen2019,Forssen2009}. Despite tremendous advances in ab-initio approaches, no unified, interpretable model currently exists that captures both global trends and local structure effects such as closed-shell gaps, odd-even staggering, shape coexistence and pairing correlations. Accurate knowledge of charge radii is essential for testing nuclear theoretical models, benchmarking advances in approaches and validating new measurements. 

The growing availability of high-precision charge radii data presents an opportunity to leverage machine learning (ML) techniques not only for accurate predictions but also for discovering underlying physical relationships. ML has recently emerged as a powerful complement to traditional nuclear models offering a flexible means to capture complex and nonlinear correlations between nuclear observables. Traditional ML models such as deep neural networks and ensemble trees have shown considerable success in predicting nuclear properties~\cite{Boehnlein2022}, also for the charge radii~\cite{Utama2016,Ma2020,Dong2023,Shang2024,Li2025}; however, their lack of transparency, and their ``black-box" nature often limits their utility in physics-driven model development. In nuclear physics, where interpretability is paramount, it is essential that ML frameworks do more than predict- they must also explain. A model that merely fits the data is of limited value unless it reveals the governing physical laws embedded in those data. A recent study~\cite{Munoz2025} addressed this issue by applying symbolic regression to reverse-engineer mathematical expressions from data with the help of available nuclear models. However, symbolic regression alone often struggles with noisy, high-dimensional nuclear data sets and requires significant efforts to yield meaningful results~\cite{Munoz2025}. 

In this work, we introduce a hybrid physics-driven ML approach that marries the high-accuracy predictive power of modern numerical regression methods with interpretable symbolic regression to extract expressions for nuclear charge radii. All features have a simple physical meaning such as mass number dependence $A^{1/3}$, proton number $Z^{1/3}$ variation, binding energy, isospin asymmetry, Casten factor, and pairing gap- an observable reflecting nucleon correlations not previously employed in ML studies of charge radii. We find that symbolic regression on data-driven outputs of optimized ML models, due to the latter's smoothened nature, is vastly more efficient than its application on raw experimental data. 

Most of the existing ML applications in nuclear physics rely heavily on a single initial random train-test split to evaluate model performance. Due to the statistical sensitivity on the random seed, this practice can lead to overly optimistic and non-reproducible results, especially when applied to the small and often skewed nuclear datasets. In standard ML applications, particularly in fields like computer vision or natural language processing, models are typically trained on datasets containing millions or even billions of elements~\cite{Pedregosa2011,Lecun1998}. Hence, our efforts should not only be focused on achieving low root-mean-squared errors but also on obtaining an unbiased and robust framework to learn about physics beyond the traditional nuclear models. Furthermore, a common limitation across ML studies of nuclear charge radii is their reduced performance for lighter nuclei, often requiring the exclusion of this region. 

\begin{figure}
    \centering
    \includegraphics[trim={2cm, 1cm, 2cm, 1cm},width=0.5\textwidth]{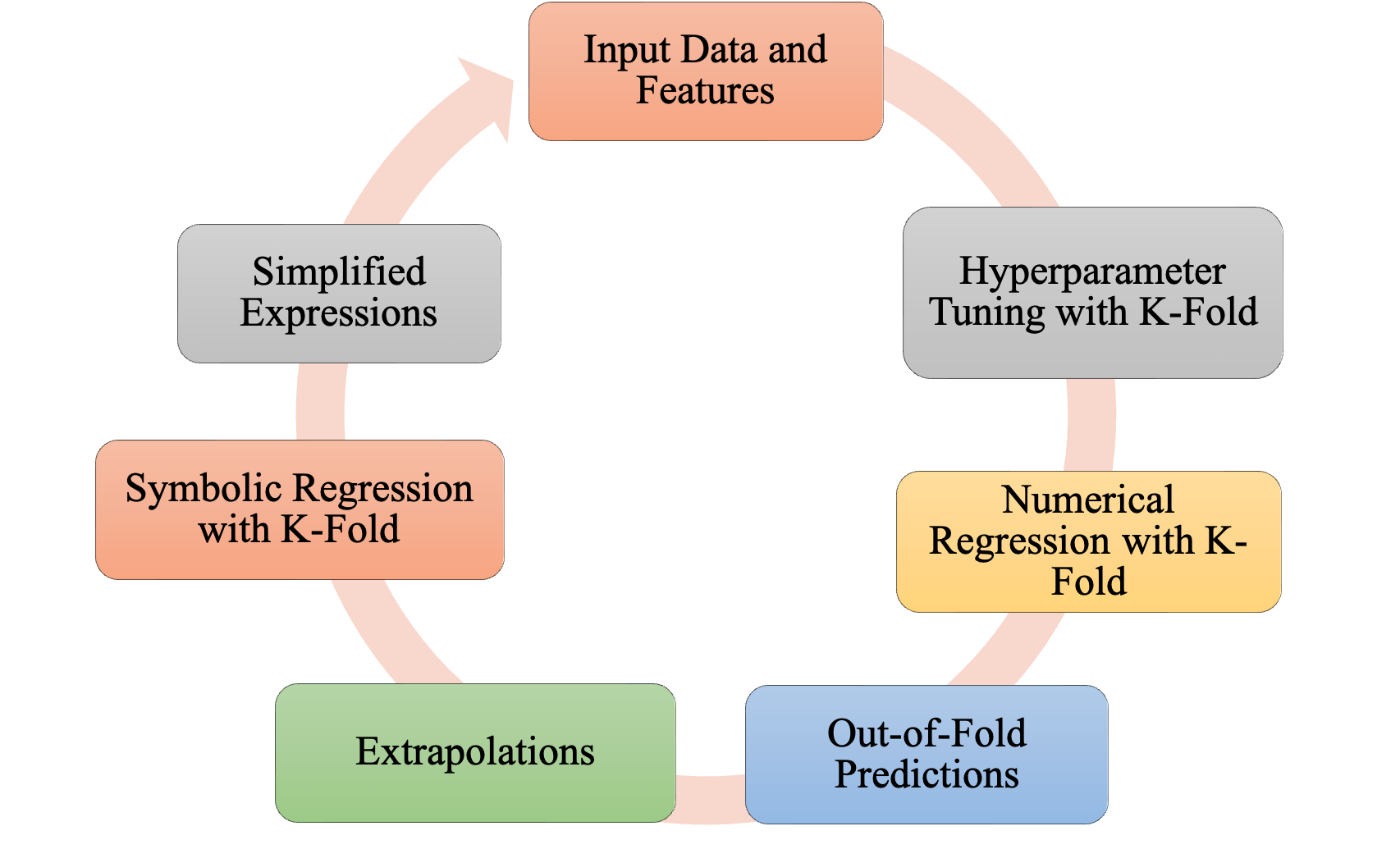}
    \caption{Cycle of a robust machine-learning algorithm from predictions-plus-extrapolations to the discovery of expressions.}
    \label{fig:ml}
\end{figure}

\begin{figure}[!htb]
    \centering
    \includegraphics[trim={0mm 0mm 0mm 0mm},width=0.32\textheight]{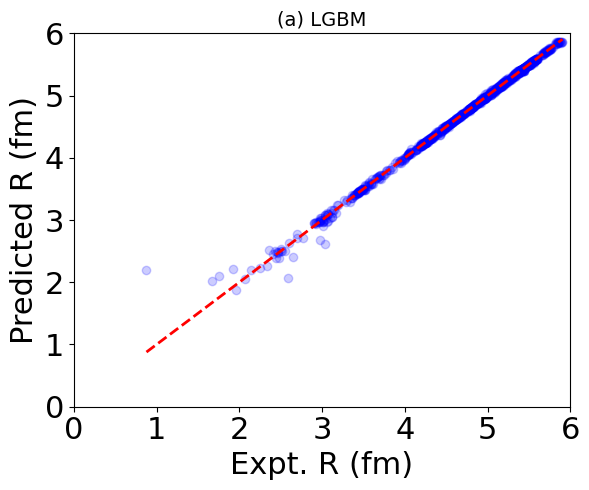}
    \includegraphics[trim={0mm 0mm 0mm 0mm},width=0.32\textheight]{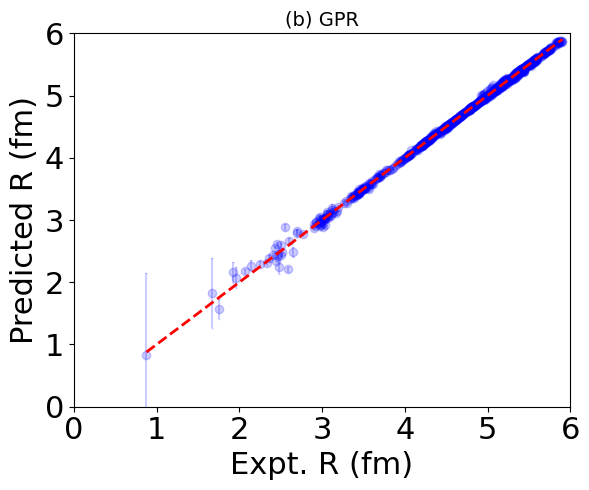}
    \caption{Machine learning (a) LGBM and (b) GPR model predicted charge radius versus experimental data. }
    \label{fig:pred}
\end{figure}

\begin{figure}[!htb]
    \centering
    \includegraphics[trim={10mm, 10mm, 0mm, 5mm},width=0.23\textwidth]{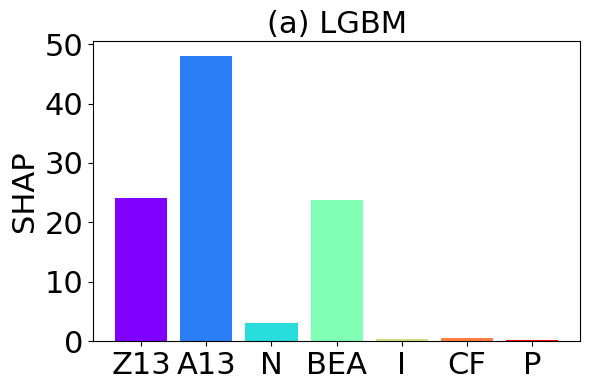}
    \includegraphics[trim={0mm, 10mm, 0mm, 5mm},width=0.24\textwidth]{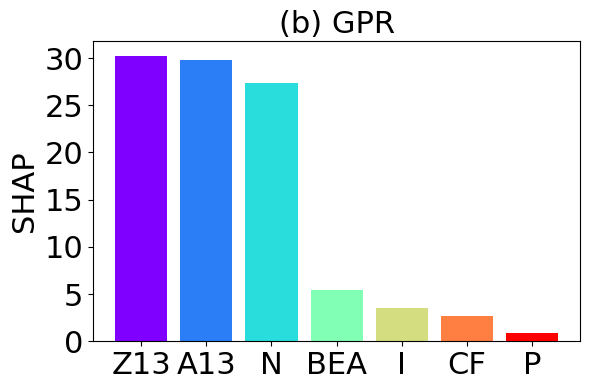}
    \caption{SHAP contributions in percentage for the features used in optimizing (a) LGBM and (b) GPR models.}
    \label{fig:shapl}
\end{figure}

\begin{figure}[!htb]
    \centering
    \includegraphics[width=0.4\textheight]{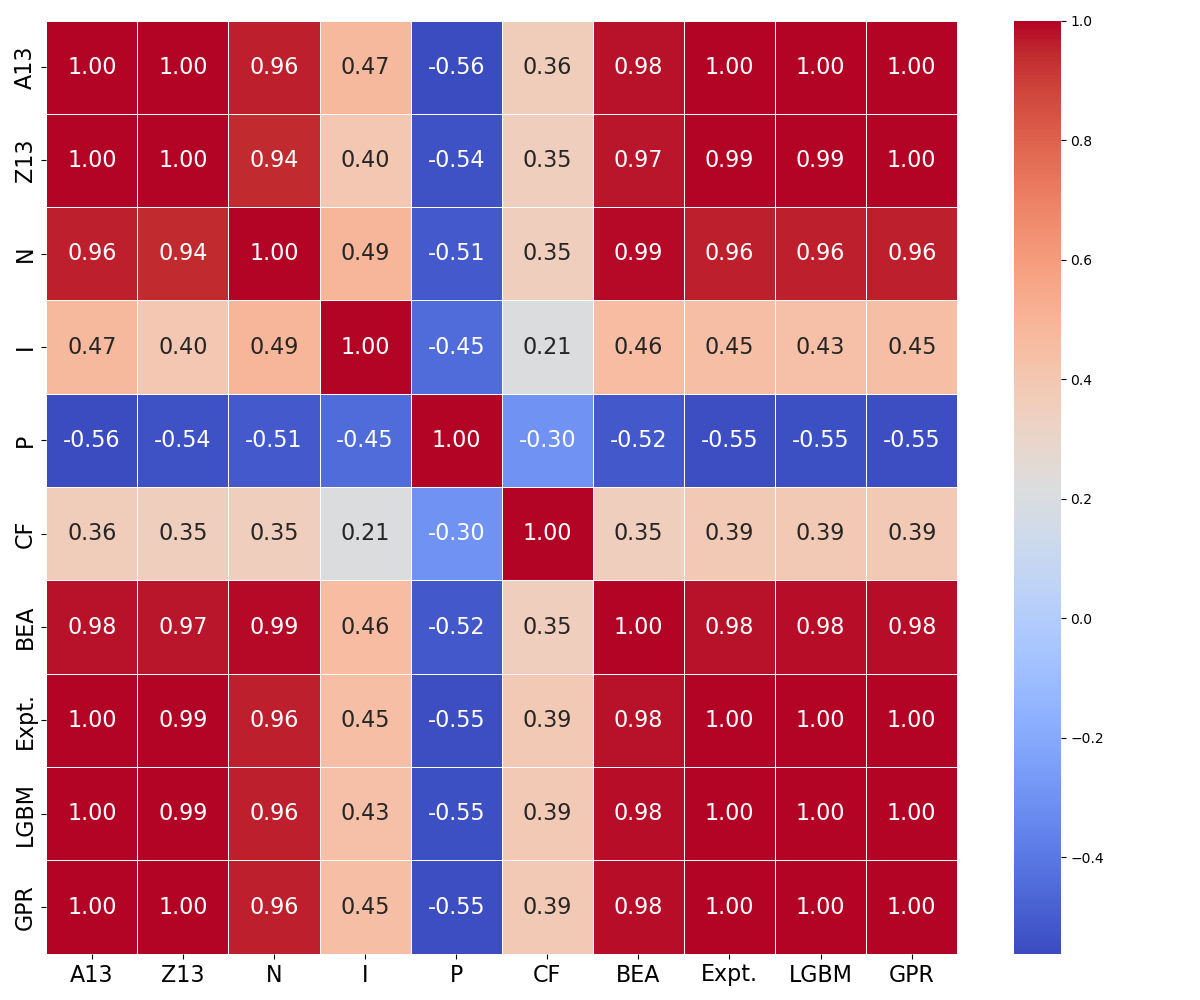}
    \caption{Correlation matrix of experimental, LGBM predicted and GPR predicted charge radii with input features.}
    \label{fig:corr}
\end{figure}

In this work, rigorous K-fold cross-validation~\cite{kfold} and automatic hyperparameter optimization ensure reproducible accuracy across the full dataset of measured nuclei, avoiding biases from single train-test splits. Applied across the full nuclear chart, from the lightest proton to heaviest curium isotopes, our framework employs the well-tested Light Gradient Boosting Machine (LGBM)~\cite{lgbm} and Gaussian Process Regression (GPR)~\cite{gpr} approaches. The K-fold cross-validation procedure ensures that each data point is treated as an out-of-fold sample exactly once, providing an unbiased set of out-of-fold predictions. Importantly, this validation extends beyond the automated hyperparameter tuning~\cite{optuna}, offering a credible assessment of the model. It also diminishes the possibility of under/overfitting to a particular train/test partition and offers a better stability of model accuracy, especially for extrapolations, important for future experimental campaigns. 

The numerically regressed LGBM and GPR results not only match state-of-the-art predictive accuracy but also serve as a foundation for discovering physics expressions using symbolic regression. Such symbolic regression naturally reproduces the $A^{1/3}$ dependence, reflecting a constant nuclear density, and introduces correction terms associated with surface and Coulomb effects. Remarkably, the derived expression mirrors the volume, Coulomb and asymmetry contributions of the liquid-drop model, emerging entirely from the data and physics-driven features rather than from pre-defined expressions of nuclear models. This work represents a step forward in physics-driven interpretable ML for nuclear physics, providing a blueprint for symbolic understanding of complex nuclear-physics phenomena through data-driven ML frameworks. The resulting framework delivers accurate global predictions while offering a pathway or interpretable, physics-informed discovery in nuclear structure.  
     
\section{Algorithm}

The ML (cyclic) algorithm presented Fig.~\ref{fig:ml} not only improves predictive performance but also uncovers latent correlations between physical properties. Our approach integrates robust-feature engineering, automated hyperparameter optimization, cross-validation, extrapolation, and model interpretability techniques. Compared to the recent nuclear-physics literature, this cycle involves additional steps of automated hyperparameter tuning, exhaustive K-fold analysis over the entire data set, followed by symbolic regression to derive analytical expressions. 
We study the variation of charge radii across nuclear chart and explore its physical correlations to other nuclear observables. Inspired by the liquid-drop model, we select features including the proton number $Z^{1/3}$ and mass number $A^{1/3}$ dependence, denoted as $Z13$ and $A13$, respectively. We augment these with total binding energy ($BEA$, in keV), neutron number ($N$), isospin asymmetry parameter $I=(N-Z)/A$, and Casten factor $CF=(VZ)(VN)/(VZ+VN)$~\cite{Casten1987} where $VZ$, $VN$ are the valence proton and neutron numbers from the nearest closed shells (2, 8, 20, 28, 50, 82, 126)~\cite{Li2025}. Crucially, we introduce the pairing gap ($P$, in keV) as a feature. Despite its known influence on nuclear structure, the pairing gap has not been utilized in previous ML studies of charge radii~\cite{Utama2016,Ma2020,Dong2023,Shang2024,Li2025} but we find that it significantly enhances the model predictions for Ca isotopes~\cite{Steppenbeck2013, Ruiz2016, Miller2019} and other lighter nuclei~\cite{Forssen2009}. 

\begin{figure}[!htb]
    \centering
    \includegraphics[width=0.33\textheight]{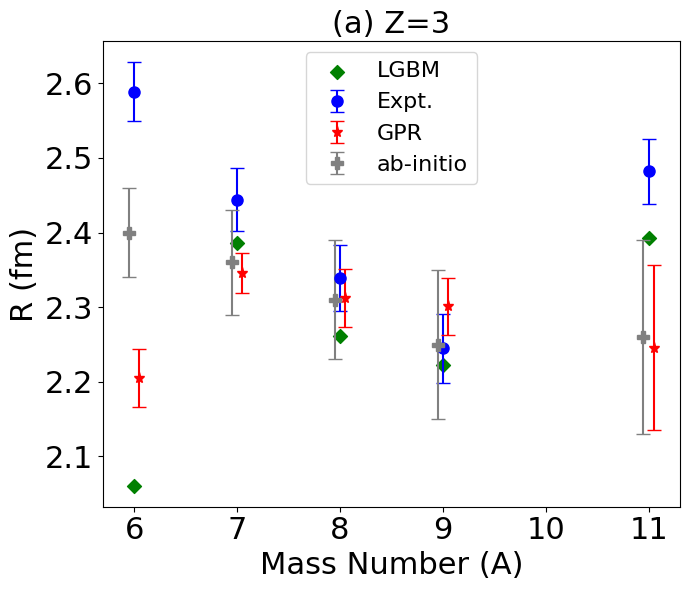}
    \includegraphics[width=0.33\textheight]{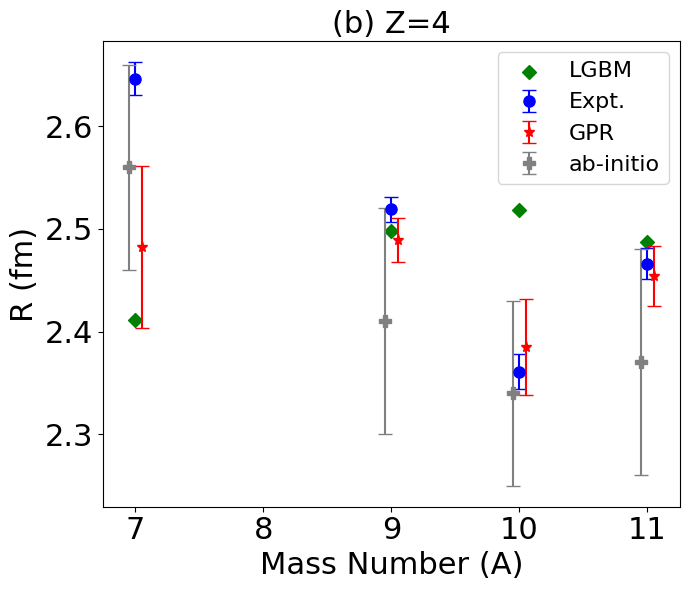}
    \caption{Measured charge radii of the Li $(Z=3)$ and Be $(Z=4)$ isotopes compared with LGBM and GPR out-of-fold predictions. Also shown are the ab-initio results~\cite{Forssen2009}.}
    \label{fig:li}
\end{figure}

\begin{figure}[!htb]
    \centering
    \includegraphics[width=0.33\textheight]{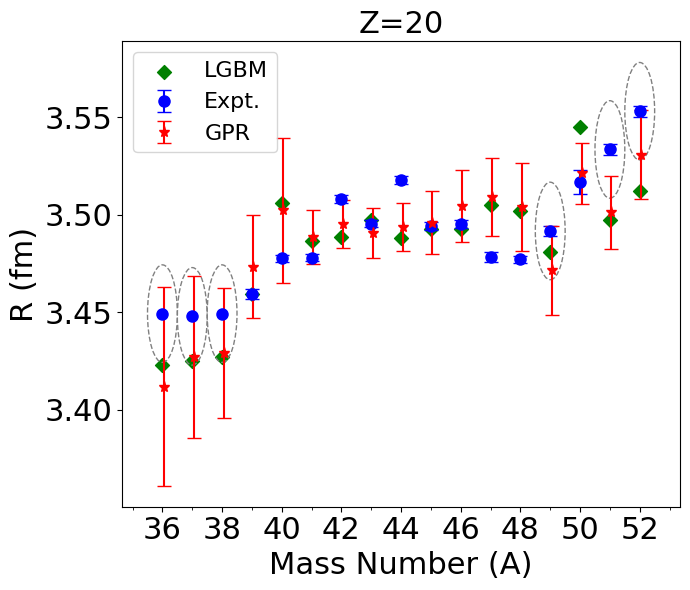}
    \includegraphics[width=0.33\textheight]{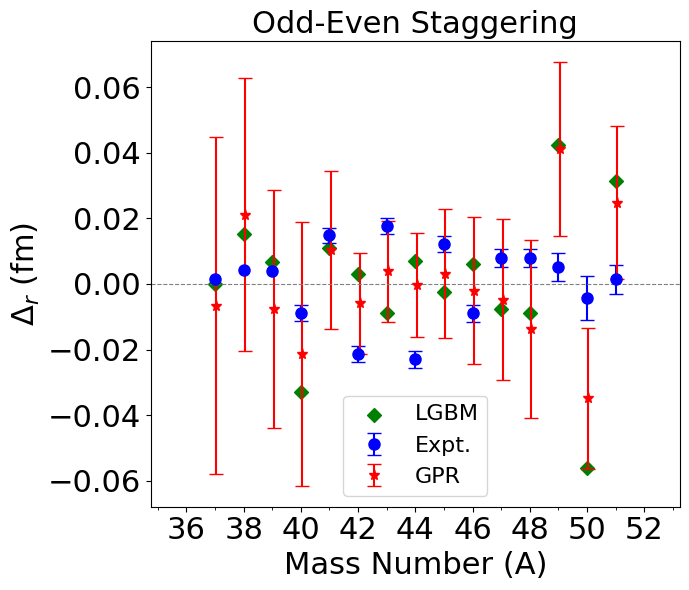}
    \caption{Measured charge radii of the Ca $(Z=20)$ isotopes compared with LGBM and GPR out-of-fold predictions. Extrapolations are highlighted with dashed ellipses. Odd-even staggering, $\Delta_r$ defined in the text, is also shown.}
    \label{fig:ca}
\end{figure}

The dataset of 956 known charge radii is adopted from reference~\cite{Munoz2025} for direct comparison. Pairing gaps and binding energies are directly taken from NNDC~\cite{nndc}. Missing pairing gaps are imputed via a mean strategy using $\mathrm{\textit{SimpleImputer}}$~\cite{scikit}. To prevent data leakage, this imputation is embedded within a pipeline that is executed independently for each fold during cross-validation. All features and the target (charge radii) are standardized using $\mathrm{\textit{StandardScaler}}$~\cite{scikit} to ensure that they contribute uniformly during model training.  

We employ two complementary ML models for comparative evaluation. LGBM, a high-performance gradient boosting framework that uses tree-based learning algorithms, has proven effective even on small nuclear datasets~\cite{Li2025}, despite its original design for large-scale data~\cite{lgbm}. GPR is a non-parameteric, kernel-based probabilistic model that provides not only point predictions but also uncertainty estimates (standard deviations) for its predictions~\cite{gpr,gprbook}, making it well-suited for small data regimes. We configure GPR with a constant multiplied by a Matern, a widely used kernel for its flexibility with $\mathrm{\textit{nu}}$ parameter controlling the smoothness and an additive $\mathrm{\textit{WhiteKernel}}$ to account for noise~\cite{gpr,gprbook}. Hyperparameters for both models are automatically optimized in 100 search iterations using the Optuna library ~\cite{optuna} within a four-fold cross-validation scheme to maximize performance; see supplementary material.

With the optimal hyperparameters identified, we proceed to train the models using an independent four-fold cross-validation. For each fold, a LGBM model and a GPR model are trained on a subset of the data $(75\%)$ and then used to make predictions on the out-of-fold validation set $(25\%)$. These out-of-fold predictions from each fold are then concatenated to form a complete set of predictions for the entire dataset, ensuring an unbiased evaluation. Model performance is quantified using root-mean-squared-error (RMSE) metrics. 

To understand the model's decisions and the relative importance of each feature, we conduct a SHAP (SHapley Additive exPlanations)~\cite{shap} analysis. For LGBM, we utilize $\mathrm{shap.TreeExplainer}$~\cite{shap} to compute the exact SHAP values. Due to the complexity of SHAP in non-parametric models, the GPR interpretation is facilitated via $\mathrm{shap.KernelExplainer}$~\cite{shap} to approximate SHAP values. The optimized models, retrained on the full dataset of 956 nuclei, are then used to extrapolate charge radii for 2601 additional nuclei where binding energies are known. 

Finally, we apply symbolic regression~\cite{pysr} to distill the intricate relationships learned by these complex black-box ML models. Symbolic regression, typically rooted in genetic programming~\cite{Poli2008}, operates by evolving a population of mathematical expressions. It begins with randomly generated formulas which are then iteratively refined across generations. Using the full dataset of 3557 nuclei (predictions-plus-extrapolations) as input, the algorithm evolves a population of mathematical expressions over 1000 iterations, minimizing the mean-squared-error $(output-input)^2/3557$ function with a four-fold cross-validation loop. This process ultimately converges on a simplified mathematical expression that approximates the learned relationships of the LGBM and GPR models. This hybrid approach is highly efficient, as it bypasses the need for an extensive physics-based guidance that is typically needed to extract meaningful formulas when applying symbolic regression directly to raw experimental data~\cite{Munoz2025}. 

\section{Results}

\begin{figure}[!htb]
    \centering
    \includegraphics[width=0.33\textheight]{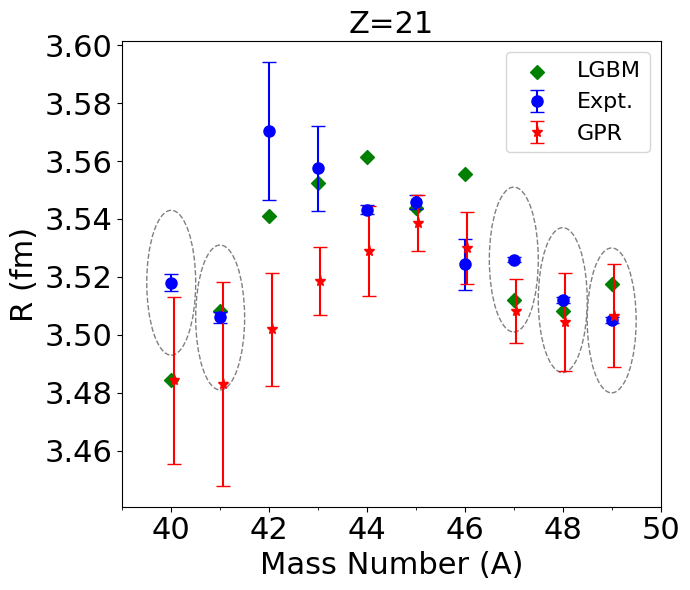}
    \includegraphics[width=0.33\textheight]{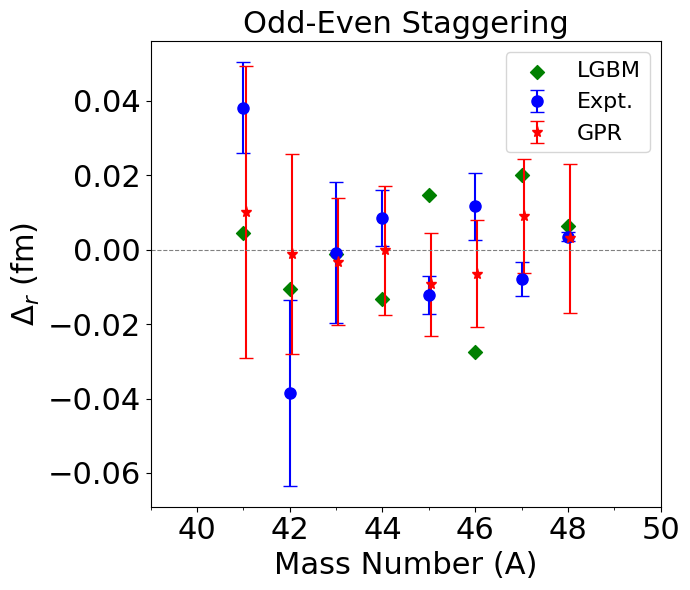}
    \caption{Measured charge radii of Sc $(Z=21)$ isotopes compared with LGBM and GPR out-of-fold predictions. Extrapolations are highlighted with dashed ellipses. Odd-even Staggering for these isotopes is also depicted. }
    \label{fig:sc}
\end{figure}

Our dataset contains 3557 nuclei with known binding energies~\cite{nndc}, of which 956 have measured charge radii adopted from ~\cite{Munoz2025}. We note that a larger compilation of over 1030 measured radii is available~\cite{Angeli2013,Li2021}, with the excluded data points providing a valuable set for an independent validation of extrapolation. The dataset of $956$ nuclei forms the basis for model training and validation via out-of-fold prediction. It spans from the proton to heavy curium $(Z=96)$ isotopes, with radii ranging from $0.87$ fm to $5.90$ fm. The remaining 2601 nuclides are used for extrapolation and are tested against the remaining measured values from the compilation~\cite{Li2021} and from recent literature~\cite{Bai2025}. 

The performance of both models evaluated on the out-of-fold predictions is excellent. The RMSE values are: for LGBM; RMSE= $0.0561$ fm, and for GPR; RMSE= $0.0315$ fm. This performance difference between the models stems from the fundamental differences in their methodologies. The non-parametric Bayesian nature of GPR allows it to better represent complex, small-scale variations in the data along with the uncertainties on the predictions, though at a somewhat higher computation cost than LGBM.  Predictions from both the models are compared with data in Fig.~\ref{fig:pred}. The red dashed line represents the perfect prediction. The ability of GPR reproducing the lighter nuclei is particularly noticeable. 

The SHAP analysis for both models, shown in Fig.~\ref{fig:shapl}, confirms the importance of mass and proton number dependencies $(A13, Z13)$, aligning with phenomenological charge radii models~\cite{Weizsacker1935,Duflo1994}. To complement SHAP, we have performed an independent correlation analysis between all input features and charge radii (experimental and LGBM, GPR predictions) as depicted in Fig.~\ref{fig:corr}, confirming and expanding upon the insights obtained so far. All chosen features are strongly correlated with the charge radius. Interestingly, the pairing gap $P$ shows a weaker SHAP contribution but exhibits a stronger correlation with charge radii than the isospin asymmetry parameter $I$ or the Casten factor $CF$, suggesting that its role may be underrepresented in the SHAP-based interpretation. The consistency of correlation patterns between experimental charge radii and the models' out-of-fold predictions further supports the interpretability and physical relevance of our framework. 

\section{Discussion}

ML models demonstrate a remarkable performance even at the lightest mass limit. The charge radius of $^{1}$H has been measured to be $0.8783 (0.0086)$ fm, where the GPR predicts $0.8293 (1.316)$ fm. Although the predicted error bar is huge, this result is nonetheless noteworthy since it concerns a proton, not even a nucleus. Both GPR and LGBM models predict reasonably well the charge radius of the deuteron as $2.1994$ fm, and $2.2624 (0.0888)$ fm, respectively compared to the experimentally measured value of $2.1421 (0.0088)$ fm. GPR also provides a reasonable estimate for the triton, $1.5741(0.1629)$ fm, while the measured value is $1.7591(0.0363)$ fm. These results are interesting considering that the model was not informed about the specific characteristics of light, halo, exotic, or bubble nuclei during training. 

The charge radii of $^{3}$He, $^4$He, $^6$He, and $^8$He have been experimentally determined to be $1.9661 (0.003)$, $1.6755 (0.0028)$, $2.066 (0.0111)$, and $1.9239 (0.0306)$ fm, respectively. The corresponding predictions from the GPR model are $2.0748 (0.1728)$, $1.8313 (0.5650)$, $2.1773 (0.0396)$, and $2.1573 (0.1654)$ fm. Considering that all predictions are discussed from the out-of-fold dataset, that is, they pertain to nuclei that were not included in training data in any fold, these results are encouraging. We further compare the results in Fig.~\ref{fig:li} for (a) $^{6,7,8,9,11}$Li and  (b) $^{7,9,10,11}$Be with the ab-initio no-core shell model~\cite{Forssen2009} and experimental data, finding excellent agreement. 
 
The calcium isotopes have garnered particular attention for their unusual charge-radius variation in both theoretical nuclear models and ML approaches. In the present work, the intriguing trend of charge radii for Ca isotopes is reasonably captured by both ML models, as shown in Fig.~\ref{fig:ca}. Extrapolations for $^{36,37,38,49,51,52}$Ca isotopes, highlighted in figure with dashed ellipses in the figure, which were not included in the 956 dataset, show good agreement with recently available data, highlighting the models' predictive power. We also show the odd-even staggering calculated using $\Delta_r=\frac{1}{2} [ R(N-1,Z)-2R(N,Z)+R(N+1,Z) ]$~\cite{Dong2023} for these isotopes. GPR catches some of the odd-even staggering between $^{40}$Ca and $^{48}$Ca but LGBM does not get it. This may be due to the missing feature $\delta=\frac{(-1)^Z+(-1)^N}{2}$. For example, Ref.~\cite{Dong2023} reproduces the odd-even staggering due to the inclusion of this $\delta$ feature. However, the odd-even staggering of charge radii is not a global phenomenon as can be seen in the data below $^{40}$Ca and above $^{46}$Ca. 

We also present the charge radius variation of scandium isotopes in Fig.~\ref{fig:sc}~\cite{Bai2025}. Only five $^{42-46}$Sc isotopes were included in the training and validation of ML models while the $^{40,41,47,48,49}$Sc isotopes are tested against extrapolations. GPR extrapolations perform particularly well, capturing the shell closure at $N=20$ and providing a reasonable estimate at $N=28$. It may most likely be linked to their binding energies. The odd-even staggering for these isotopes has also been calculated and shown in Fig.~\ref{fig:sc}.        

\begin{figure*}
    \centering
    \includegraphics[trim={1cm, 0cm, 0.2cm, 0.5cm} ,width=0.35\textheight]{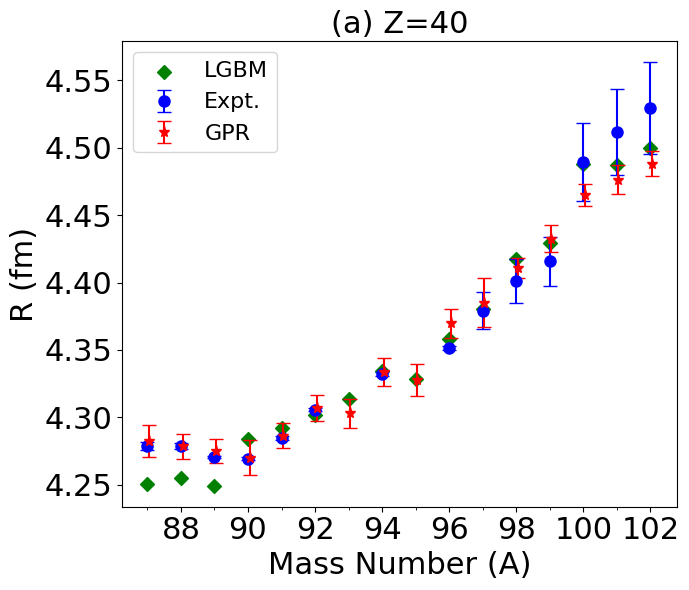}
    \includegraphics[trim={0.2cm, 0cm, 1cm, 0.5cm} ,width=0.35\textheight]{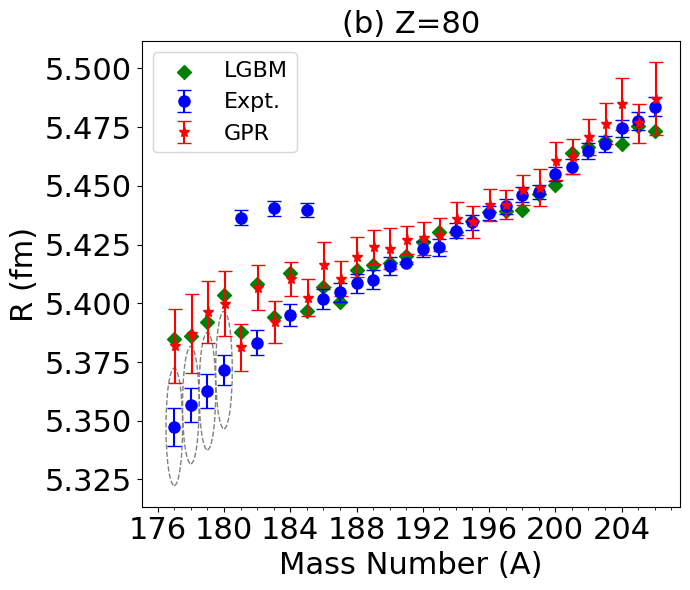}
    \includegraphics[trim={1cm, 0cm, 0.2cm, 0.2cm},width=0.35\textheight]{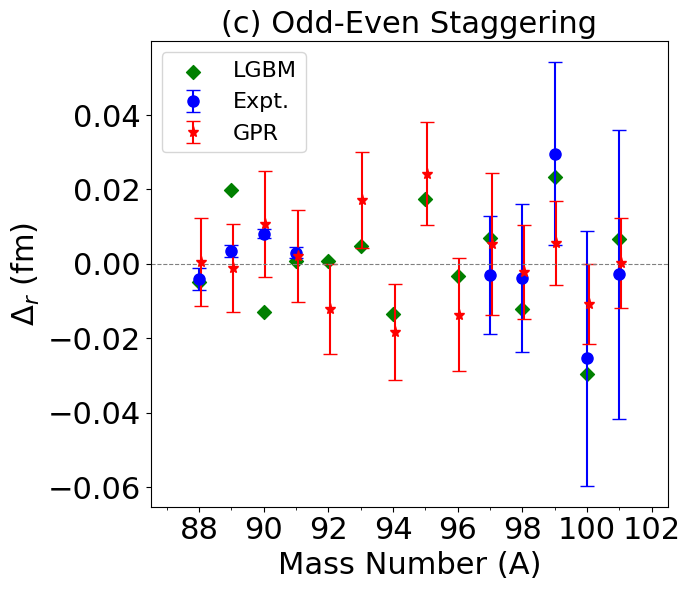}
    \includegraphics[trim={0.2cm, 0cm, 0.2cm, 1.2cm},width=0.35\textheight]{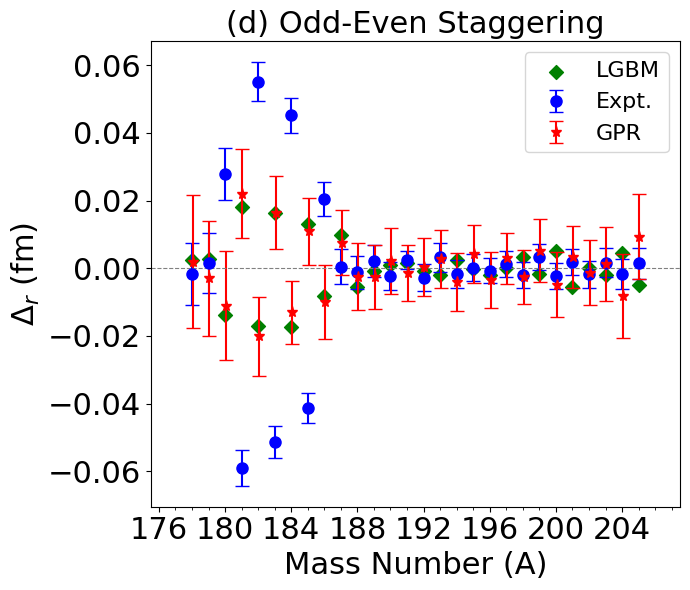}
    \caption{Experimental data~\cite{Munoz2025}, LGBM and GPR predicted charge radii of (a) Zr, $Z=40$ and (b) Hg, $Z=80$ isotopes. ML models struggle to predict the charge radii for those nuclei known to have shape coexistence beyond $A=99$ in Zr and below $A=186$ in Hg isotopes. Extrapolations are shown with dashed ellipses. Also, shown are the odd-even staggering results for (c) Zr, $Z=40$, and (d) Hg, $Z=80$ isotopes. }
    \label{fig:zr-hg}
\end{figure*}

The ML models also perform well in regions of complex structure. In Zr isotopes in Fig.~\ref{fig:zr-hg}(a), the sudden onset of deformation beyond $N=59$ is partially captured. However, in the neutron-deficient $^{181-185}$Hg isotopes shown in Fig.~\ref{fig:zr-hg}(b), a well-known region of shape coexistence, the models cannot reproduce the dramatic odd-even staggering. This indicates that while our current features capture global trends, describing rapid structural changes may require additional physics input. Besides, we also study the odd-even staggering in Fig.~\ref{fig:zr-hg}(c) for Zr isotopes and in Fig.~\ref{fig:zr-hg}
(d) for Hg isotopes. Both the ML models work reasonably well with a superior performance of GPR except the shape coexistence region of Hg isotopes. 

For the long isotopic chains of Sn and Pb, both models reproduce the data accurately, including the characteristic kink at $N=82$ and $N=126$ shell closure, respectively. The extrapolated GPR value 4.7253(0.0329) fm for $^{134}$Sn, is in excellent agreement with the experimental value of 4.7317(0.0014) fm. 

We provide the full LGBM and GPR data sets along with the experimental data in separate file. This could be useful for future experimental campaigns and theoretical modeling. 

\section{Simplified Expression}

We apply symbolic regression to the full set of 3557 predicted-plus-extrapolated charge radii from the optimized GPR and LGBM models. Using $\textit{PySRRegressor}$, we evolve expressions over 1000 iterations up to a complexity level of 15 to find the optimal balance between accuracy and simplicity. We also employ a thorough four-fold cross-validation to provide a reliable, unbiased estimate of the predictive performance of symbolic regression on unseen data before performing the final symbolic regression on the full data set. More details and the evolved expressions are provided in the supplementary material. Both models initially rediscover the simple linear relationship with $A^{1/3}$ as the most dominant feature, thereby validating the symbolic regression approach. As complexity is allowed to increase, however, both models diverge, finding distinct and physically insightful pathways to higher accuracy. 
For example, the symbolic regression of the GPR results finds an optimal balance of simplicity and accuracy at the complexity level of 8 yielding the formula (with $BEA$ in GeV)
\begin{equation}
R_{GPR}^{(8)}=0.386A^{1/3} + 0.536Z^{1/3} + 0.185BEA + 0.617,
\label{eq:eq1}
\end{equation}
with an RMSE of $0.079$ fm. The first term is the volume term of the classic liquid-drop model, supporting a constant nuclear density. The second term 
relates to the Coulomb-energy correction due to charge distribution competing with it. The third term represents a small but significant correction due to the total binding energy, $BEA$. The fourth term merely adds a correction for the missing correlations. On the other hand, the algorithm applied to the LGBM results finds the following expression at the complexity level of 8 (with $BEA$ in GeV),
\begin{equation}
R_{LGBM}^{(8)}= 0.540 A^{1/3} -0.013N + 1.655BEA + 1.265 
\label{eq:eq2}
\end{equation}
with an RMSE of $0.070$ fm. The first term in this expression also corresponds to the classic liquid-drop model. The second term is different if compared to the Eq.~(\ref{eq:eq1}), it tries to compensate the linear increase with $A^{1/3}$ using the neutron number. The third term again shows the importance of total binding energy while the last term is a constant. All these terms are physically known to be governing the underlying nuclear charge radii.  Though these expressions approximate the numerically regressed results globally with a reasonable RMSE, they are too simple to capture more localized effects, such as the characteristic kinks in isotopic chains at shell gaps. The most accurate distillation of the LGBM model at the complexity level of 15 (with RMSE of $\sim$0.05 fm) reveals a quadratic dependence on neutron number, $N$, denoted by $R_N$ in the below expression, 
\begin{equation}
R_{LGBM}^{(15)}=-0.881 R_N \times R_{AZ} + 4.756 
\end{equation}
where $R_N= (0.036(0.023N - 1.889)^2 - 0.754)$ and $R_{AZ}= (0.972A^{1/3} + 0.647Z^{1/3} - 7.281)$. The non-linear term $R_N$ is capable of modeling a changing slope in an isotopic chain, giving rise to a subtle kink at the shell gap, which is absent at the complexity level of 8. The $R_{AZ}$ term is similar to what algorithm has obtained at the simpler complexity level depicting the major contributions of $A^{1/3}$, and $Z^{1/3}$ dependence on charge radii. The most accurate GPR-distilled expression achieves its low RMSE through a completely different mechanism, by designing a multi-variable interaction term $R_I$, linking $BEA, CF$ and $A^{1/3}$,
\begin{equation}
R_{GPR}^{(15)} = 0.026 CF + 0.728 I + 1.202 Z^{1/3} + 0.063 
- 0.062 R_I 
\end{equation}
where $R_I=(1.907BEA - 2.118)
((0.972A^{1/3} - 4.862)(0.320CF - 1.140) - 0.435)$. This expression exhibits the role of pairing correlations Casten factor, isospin asymmetry in addition to the $A^{1/3}$ and $Z^{1/3}$ dependence on charge radii. This comparison demonstrates that symbolic model distillation can not only discover simple physical expressions but also uncover the distinct, underlying physical mechanisms learned by different numerical ML regression models. Results from these expressions for both LGBM and GPR models are in good agreement with both the experimental data and the numerical regression, as shown in the supplementary material.

\section{Conclusion}

We have developed a robust and interpretable machine learning framework for nuclear physics, successfully applying it to model charge radii from the proton to heavy nuclei. By enforcing rigorous cross-validation and automated hyperparameter tuning, our approach ensures model generalizability, a critical step for the sparse and skewed datasets typical in our field. Our results demonstrate that Gaussian Process Regression is particularly well suited for nuclear physics applications, offering superior accuracy for light nuclei and providing essential uncertainty quantification. 

Both GPR and LGBM models successfully capture key nuclear structure phenomena, including shell closures and to some extent, odd-even staggering. The inclusion of the pairing gap as a feature proved valuable for improving predictions. The models' extrapolations to unknown regions show promising agreement with the data (not included in developing ML models), underscoring their predictive capability. A data file for both LGBM and GPR results is provided for the completeness.   

The capstone of our work is the successful application of symbolic regression to distill the complex ML models into a simple, interpretable mathematical formula. The formula relies on dominating features, $A^{1/3}$, or $Z^{1/3}$, and provides a direct, interpretable link between the input features and the predicted charge radius, offering physical insights derived from the data-driven numerically regressed models. However, both GPR and LGBM model results leads to different expressions owing to their different mechanisms. The importance of shell effects and binding energies is also visible required for the local nuclear structure effects especially at the higher complexities. This hybrid numerical-symbolic regression approach provides a powerful blueprint for data-driven discovery in nuclear physics and beyond. Future work will extend this framework to other observables such as magnetic and quadrupole moments, decay $Q-$values, nuclear interactions and excitations along with multi-objective analysis. 

\section*{Acknowledgments}
The author BM gratefully acknowledges the financial support from the HORIZON-MSCA-2023-PF-01 project, ISOON, under grant number 101150471. We thank Adnan Ghribi, GANIL for reading the manuscript carefully and useful suggestions.

\appendix*

\begin{small}
\begin{verbatim}

\end{verbatim}   
\end{small}

\newpage

\end{document}